# Relaxation of strongly coupled electron and phonon fields after photoemission and high-energy part of ARPES spectra of cuprates

A. E. Myasnikova, E. A. Gileeva and D. V. Moseykin

Physics Faculty, Southern Federal University, Rostov-on-Don, Russia

We consider photoemission from cuprate high-temperature superconductors taking into account joint relaxation of strongly coupled fields: a field of correlated electrons and phonon field. Such relaxation occurs in systems with predominantly long-range electron-phonon interaction at sufficiently high carrier concentration due to coexistence of autolocalized and delocalized carriers. A simple method to calculate analytically a high-energy part of the ARPES spectrum arising is proposed. The approach suggested yields all the high-energy spectral features like broad bands of Gaussian shape and "vertical dispersion" patterns being in good quantitative agreement with the experiments at any doping with both types of carriers. Demonstrated coexistence of autolocalized and delocalized carriers in cuprate superconductors is important for understanding their transport and magnetic properties. Presence of large-radius autolocalized carriers shown may be helpful in explanation of charge ordering in doped cuprates.

1. Introduction

Thirty years of theoretical modeling of cuprates demonstrating high-temperature superconductivity by separating one "main" interaction failed to explain their properties. Obviously this points out that simultaneous interplay of charge, spin and lattice degrees of freedom is crucial. Recently such complex approach was successfully applied to quasi-1D cuprates: experimental results on resonant inelastic X-ray scattering turned out to be in complete accordance with the theoretical description based on simultaneous consideration of electron relaxation and electron-phonon interaction (EPI) [1].

Here we consider photoemission from cuprate high-temperature superconductors taking into account *joint relaxation of strongly coupled fields: a field of correlated electrons and phonon field*. Broad bands with Gaussian shape [2-5] and "vertical dispersion" patterns [6-10] were observed in ARPES spectra of cuprates on both sides of cuprates' phase diagram (hole- and electron-doped) as well as in undoped parent compounds [11]. Progress in qualitative theoretical description of broad bands in ARPES spectra of cuprates was achieved due to consideration of joint influence of electron correlations (in the frames of t-J model) and strong Holstein (short-range) EPI resulting in small polaron (SP) formation [12,13]. However, "vertical dispersion" patterns have not been obtained in this approach. Attempts to obtain them in the frames of Hubbard model or other models without strong EPI [7,9,10] yielded the low-energy part of the ARPES band in good agreement with the experiments at essential doping but did not yield the broad band of Gaussian shape observed experimentally well below the Fermi energy at low doping [2-5,11].

Simultaneous presence of broad Gaussian bands and "vertical dispersion" emerges in a model with strong long-range (Frohlich) EPI [14] favoring large polarons (LP) [15]. But to obtain the ARPES spectrum features being in quantitative agreement with the experiments on cuprates simultaneous taking into account carriers correlations and strong Frohlich EPI is necessary. Therefore here we develop an approach joining results of Hubbard or t-J model with the model of strong long-range EPI. Applying the approach suggested we consider the system relaxation after photoemission (it demonstrates a large diversity of relaxation pathways for different doping and carrier momentum) and calculate the spectral function taking advantages of using the coherent states basis for the phonon field [16].

The main difference of systems with strong Frohlich EPI is coexistence of autolocalized and delocalized carriers occurring at sufficiently high carrier concentration [14]. As a result translational degeneracy of the ground state characteristic for the autolocalized state (AS) of the carrier [17] is removed and electron subsystem participates in the relaxation after photoemission. The relaxation pathways are different at photoemission from AS and from delocalized state (DS), and according to Pauli exclusion rule ASs and DSs occupy different regions in the momentum space [14]. Therefore the coexistence of autolocalized and delocalized carriers displays itself in ARPES spectrum as presence of "vertical dispersion" universally observed in superconducting cuprates [6-10]. Reconstruction of the phonon vacuum during relaxation when initial or final carrier state or both these states are autolocalized results in broad ARPES bands of Gaussian shape [16]. Thus, generalizing the relaxational approach [13,16,18-21] to higher doping case, here we develop the approach that allows considering photoemission at arbitrary carrier concentration in a system with strong electron correlations and strong long-range EPI.

The model under consideration allows simple analytical calculation of high-energy part (HEP) of the ARPES spectrum. The suggested method takes advantages of two facts. First, the phonon vacuum reconstruction during the system relaxation is accompanied with radiation of different number of phonons in different acts with the Poissonian distribution of the probability, the average number of phonons is determined by the change of the phonon vacuum energy [16, 20]. Second, the energy conservation equation rigidly ties the number of radiated phonons with the binding energy of photoelectron.

Only two system characteristics are necessary for the calculation: dispersion of the correlated carriers and effective dielectric constant $1/\varepsilon^* = 1/\varepsilon_\infty - 1/\varepsilon_0$ characterizing the strength of the long-range EPI [22]. The former can be taken either from theoretical models (for example, t-J model [23]) or extracted from ARPES spectrum of undoped or highly doped cuprates because in a certain area of the momentum space they follow the "bare" carrier dispersion (shifted deeper in the undoped case) as it is shown below. The latter, in principle, can be calculated using static and high-frequency dielectric constants. However, such approach would neglect renormalization of the EPI strength due to its interplay with other important interactions [24]. Therefore here we use already renormalized effective dielectric constant deduced from ARPES spectrum of undoped or low-doped cuprates.

Predicted energetic and momentum position and line-width of the features in the HEP of ARPES spectrum of cuprates are in good *quantitative* agreement with the experiments at any doping with both types of carriers [2-11]. In particular, the double scale of energy of high-energy anomaly (HEA) in electron doped cuprates [9,10] with respect to that in hole doped ones [6-8] receives natural explanation. Thus, combination of two models - Hubbard or t-J model with the model of strong Frohlich EPI - generates new approach that results in significantly improved agreement of calculated HEP of ARPES spectrum of cuprates with experiments. The approach suggested modifies our notion about the ground state of cuprates by demonstrating coexistence of autolocalized and delocalized carriers. In future works this finding may help fitting better temperature and doping dependence of other characteristics of cuprates (e.g. transport, optical and magnetic). The results obtained allow us also making some notes on such long discussed problems as appearance/disappearance of HEA in some cuts in different BZs at photon energy change [25-27] (sometimes considered as confirmation of matrix-elements nature of HEA) and charge ordering in doped cuprates [28-31].

The article is organized as following. First we discuss the model that allows considering high carrier concentrations in a system with simultaneous presence of strong electron correlations and strong long-range EPI. Then we briefly discuss the variety of the relaxation pathways in such a system after its photoexcitation. Next section describes suggested analytical method to calculate the HEP of ARPES spectrum from such systems. Then it is applied to systems doped with electrons and holes to any doping level including undoped system. The calculated spectra are compared with the experimental ARPES spectra of cuprates from undoped up to overdoped with electrons or holes, good quantitative agreement is demonstrated. In conclusion we briefly discuss the implications of the results obtained.

2. A model of system with strong electron correlations and strong long-range EPI

LP is formed by a carrier at strong long-range EPI, if the bandwidth $W$ of the "bare" carrier exceeds its average kinetic energy in the polaron [15, 32] which is equal to the polaron binding energy $E_p$ [22]. In the opposite case as well as at the dominance of short-range EPI the SP is formed [15, 32]. For the Lower Hubbard Band (LHB) both $W$ and $E_p$ values can be extracted from the dispersion of ARPES band at zero or low doping with holes. Indeed, it follows the "bare" carrier dispersion [12,13] shifted deeper due to hole polaron formation, and binding energy $\varepsilon_{max}$ in its maximum is about $2E_p$ for SP [32] and $3E_p$ [22], or, more precisely, about $3.2E_p$ for LP [16]. The polaron binding energy $E_p$ in the Upper Hubbard Band (UHB) is obtained similarly from the position of the ARPES band maximum at low doping with electrons [4,5]. The UHB bandwidth can be deduced from the spectrum at high doping with electrons in which the dispersion of the ARPES band crossing the Fermi level also follows "bare" carrier dispersion as is shown below. Since in cuprates $W \approx 0.4 \div 0.5$ eV for the LHB [11] and is larger for the UHB [9], whereas $|\varepsilon_{max}| \approx 0.42$ eV [4,5,11] in both LHB and UHB, LP model is appropriate for both bands. This conclusion is confirmed by comparison of the predictions obtained below with the experiments.

Carrier tunneling between nodes inside the LP polarization potential well is much quicker (adiabatical) than the ions motion, as ratio of the carrier average kinetic energy in LP $E_{kin}=E_p \approx \varepsilon_{max}/3.2$ [16] to the average phonon energy [13] shows. Therefore electron correlations can be taken into account in the effective mass approximation. The dispersion of correlated carriers can be taken either from theoretical calculations [23] or from experimental ARPES spectra of cuprates [2-11] (excluding the region separated by "vertical dispersion"), the more so that they demonstrate good correspondence. Then we consider autolocalization of carriers with the momentums near extremes of the correlated carrier bands. Resulting dispersion of UHB and LHB modified by strong Frohlich EPI are shown by Fig.1a,b, respectively. In these figures and in further calculations we use the following dispersion for nodal direction ($k_x=k_y$) in UHB and LHB (in eV)

$$E_{UHB}^{nodal}(\mathbf{k}) = -0.5\big(\cos(k_x a) + \cos(k_y a)\big) + 1, \tag{1}$$

$$E_{LHB}^{nodal}(\mathbf{k}) = -0.125\big(\cos(2k_x a) + \cos(2k_y a)\big) - 0.25, \tag{2}$$

where $a$ is the lattice constant; zero energy in (1) and in (2) is in the bottom of UHB and top of LHB, respectively, that corresponds to very low doping with electrons (Eq.(1)) or holes (Eq.(2)). The UHB dispersion is taken from the experimental ARPES spectrum of electron-doped cuprate (near optimal doping) [9] in nodal direction, this possibility is substantiated below. The LHB dispersion corresponds both to experimental ARPES spectrum of undoped cuprates [11] shifted upward by $|\varepsilon_{max}|$ and theoretical t-J model dispersion [23] in nodal direction as they are in good agreement with each other. The reasons of the agreement and the shift are also discussed below.

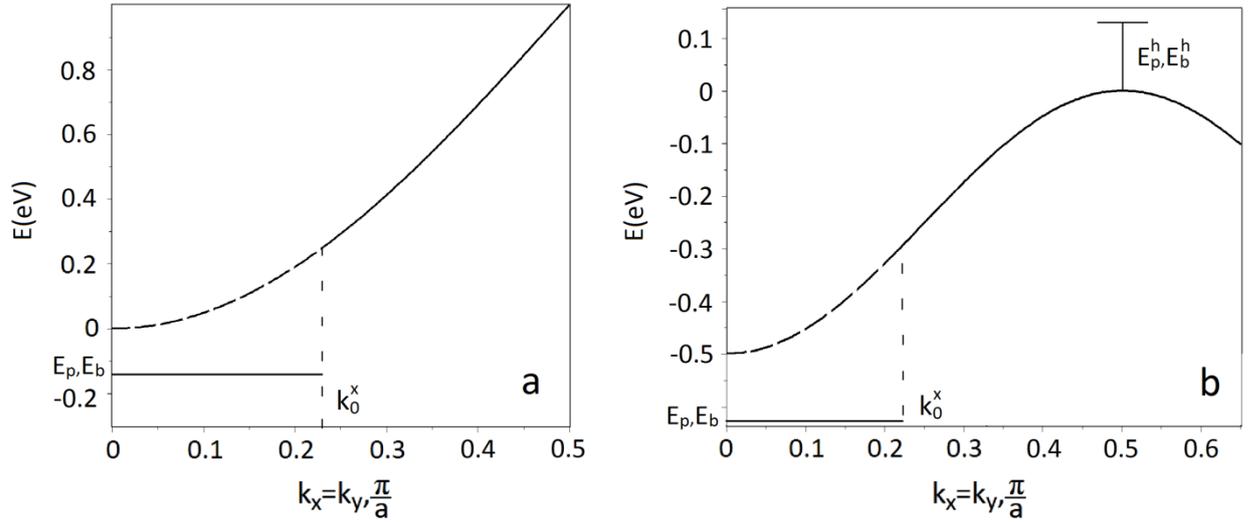

Fig.1a, b - Upper and Lower Hubbard Band dispersion in nodal direction (expressed by Eqs.(1,2)) modified by strong Frohlich EPI, respectively, $k_0^x$ is $k_0$ projection on x or y axis, $E_p^h, E_b^h$ stand for binding energy of hole polaron and bipolaron.

Besides carrier dispersion, the only value we need to calculate the HEP of the ARPES spectrum from system with any doping level is effective dielectric constant $\varepsilon^*$. As is described in the Introduction it should be deduced from experimental ARPES spectrum using relation between the binding energy $\varepsilon_{max}$ in the ARPES band maximum and the polaron binding energy $E_p$: $E_p \cong \varepsilon_{max}/3.2$ [16,22] and the well-known Pekar result [22] (in eV):

$$E_p = 1.47(\varepsilon^*)^{-2} m^*/m_e. \qquad (3)$$

Eq.(3) is obtained for the isotropic case but contains the carrier effective mass which is anisotropic in the considered cuprates. Nevertheless, we can use Eq.(3) with "effective" isotropic $m^*$ to determine "effective" $\varepsilon^*$ without introducing essential mistake into further calculations as the value of $\varepsilon^*$ enters in them only in expressions for energies (bipolaron binding energy and the phonon field energy in the bipolaron, Eqs.(12) below) which contain $\varepsilon^*$ and $m^*$ similarly to Eq.(3). Therefore we use below the carrier effective mass $m^* \approx m_e$ near the bands extremes according to Eqs.(1,2). Then Eq.(3) yields $(\varepsilon^*)^{-1} \approx 0.3$ for both electron- and hole-doped cuprates and we use this value in further calculations.

The carrier momentum in LP and in large bipolaron has large uncertainty $\hbar k_0$ tied with the polaron volume $V_0$ (the large polaron and bipolaron have close volumes [14]) by the uncertainty relation:

$$\frac{4}{3}\pi(\hbar k_0)^3 V_0 = (2\pi\hbar)^3. \qquad (4)$$

$k_0$ is ordinarily about $\pi/(4a) \div \pi/(3a)$, therefore momentum is not suitable to characterize the carrier AS. Spontaneous breaking of the translational symmetry in a system with strong EPI results in the ground state degeneration as polarons can be located in different regions of the crystal [16]. Thus, ASs can be characterized by their center position, different carrier states take place when the distance between the AS centers is larger than the AS diameter. The binding energy per carrier in large bipolaron calculated with taking into account carrier correlation is close to that in LP [14], therefore the ground state of the system with strong Frohlich EPI at low doping can be either polarons or bipolarons. At higher carrier concentrations it is a system of bipolarons [14].

The maximum AS momentum is much smaller than $\hbar k_0$ because phonon group velocity limits the AS velocity [33]. Hence there exists a maximum concentration of carriers in ASs which is roughly $n_0 = 2/V_0$, where $V_0$ is the AS volume. At the carrier concentration $n > n_0$ and sufficiently low temperatures delocalized carrier states with $|\mathbf{k}| \equiv k < k_0$ are prohibited by Pauli exclusion rule. Therefore they are shown with dashed line on Fig.1a,b. At high carrier concentration the system ground state corresponds to essentially smaller bipolaron radius than that at low doping. For example, at $(\varepsilon^*)^{-1}=0.3$, $m^* = m_e$ and $\varepsilon_\infty = 3$ [34] the single bipolaron radius $R_{bip} \approx 13.5$ Å [14] (we calculate it as the radius of the region, that contains 0.9 part of the polarization charge), whereas at $n \geq n_0$ it can be estimated as $R_{bip} \approx 6.5 \div 7$ Å. According to Eq.(4) the corresponding $k_0$ value is about 0.35-0.37Å$^{-1}$. When the momentum is along the nodal direction the corresponding value of $\mathbf{k_0}$ projection on $k_x$ or $k_y$ axis is $0.2 \div 0.27$ Å$^{-1}$, where different possible values of the momentum projection onto the normal to the crystal surface are taken into account.

Pauli exclusion rule also causes absence of EPI screening by delocalized carriers. In undoped parent compounds screening is absent due to completely filled band. At carriers concentration $n<n_0$ and low temperatures there are no delocalized carriers. At $n>n_0$ delocalized carriers are present in the system. But to form screening charge the squared modulus of a DS wave function should be non-zero in a localized area of the order of the polaron size. Then the carrier momentum should be from the interval $k < k_0$, which is prohibited by Pauli exclusion rule.

Ground state of a single hole at strong EPI is a hole polaron. However, at strong EPI completely filled band contains about $n_0/2$ electron bipolarons. Then where in the coordinate space do the hole ASs can locate? Obviously the hole is not localized inside the electron bipolaron as it is not profitable energetically. The more so that between electron bipolarons there are areas of negative polarization charge attracting the holes. It is worth noting that due to large radius of the carrier ASs in cuprates the electron bipolarons and the hole ASs contain phonon vacuum deformation in mainly different harmonics with the wave vectors around Γ and (π/2, π/2) points of the first Brillouin zone, respectively.

3. Variety of the relaxation pathways in a system with strong Frohlich EPI and sufficiently high carrier concentration

Taking into account the phonon subsystem relaxation after electron photoexcitation resulted some time ago in understanding the nature of broad bands in optical and ARPES spectra of cuprates. It is multiple phonon generation at creation/decay of the phonon vacuum deformation, the number of phonons is different in different acts [2,3,11-14,16,20]. In a single carrier limit the effect is similar in SP and LP models with only quantitative difference. However for systems with higher doping the models yield qualitatively different predictions as the LP case allows coexistence of autolocalized and delocalized carrier states.

Here we show that for such systems (with strong Frohlich EPI) generalizing the relaxational approach [13,16,18-21] to higher doping case allows to calculate the HEP of ARPES spectra of cuprates being in quantitative agreement with experiments at any doping with both types of carriers. The central point is participation of electron subsystem in the relaxation at $n>n_0$ as the ground state translational degeneration occurs in a system till it contains only ASs and is absent when delocalized states appear at $n>n_0$. If photoelectron originates from bipolaron state so that its in-plane momentum $k<k_0$ (supposing the band minimum is in $k=0$ point), the intermediate state with one carrier remained in bipolaronic polarization potential well (it does not change during electron phototransition according to Frank-Condon principle) in presence of delocalized carriers relaxes due to EPI in other way than in their absence. Namely, relaxation results in restoration of the bipolaron and emptying the state from the Fermi surface (or hole appearance at the Fermi surface).

If photoelectron comes from delocalized state, i.e. its in-plane momentum $k > k_0$, the system relaxation is absolutely different and simple, without changing the phonon vacuum. A "boundary" $k=k_0$ between the two types of the system relaxation with different relaxation energy displays itself in ARPES spectrum as "vertical dispersion". It should be noted that in systems where SPs are formed there is not a coexistence of localized and delocalized carriers and correspondingly there is not "vertical dispersion" pattern in the ARPES spectrum.

Even more interesting relaxation occurs in undoped or hole-doped systems. The hole left after the photoelectron escape relaxes to the Fermi surface and forms AS due to strong EPI. Such final state occurs for doping lower than the ASs maximum concentration $n_0$. At higher doping the final state is delocalized hole at the Fermi surface (as in the electron doping case with $n>n_0$). If the photoelectron originates from the electron AS (bipolaron) it is restored in the final state. Each relaxation way is discussed in details below at calculating its display in the ARPES spectrum.

4. Simple analytical method to calculate the high-energy part of ARPES spectrum from systems with strong long-range EPI

The method of calculating ARPES spectrum from systems with strong Frohlich EPI takes advantages of using the coherent states basis for the phonon field state in the (bi)polaron [33,20,16,14]. Let us consider first the photoemission from the polaron state and then (in the next sections) generalize the method to the bipolaron case and to undoped and hole doped systems. The simplest analytical calculation of the band in ARPES spectrum caused by photodissociation of the polaron is based on Fermi golden rule

$$W_{if} = \frac{2\pi}{\hbar} \left| \langle f | \hat{H}_{int} | i \rangle \right|^2 \delta(E_i - E_f) \tag{5}$$

$$\langle f | \hat{H}_{int} | i \rangle \propto \int d\mathbf{r} \exp(-i\mathbf{k}\mathbf{r}) \hat{H}_{int} \psi(\mathbf{r}) \prod_{\mathbf{q}} \langle v_{\mathbf{q}} | d_{\mathbf{q}} \rangle, \tag{6}$$

with the vector of the system initial state written in the adiabatic approximation as a product of the electron wave function $\psi(\mathbf{r})$ in the polaron and a vector of the phonon field state $\prod_{\mathbf{q}} | d_{\mathbf{q}} \rangle$ in the coherent states representation. Parameters $d_{\mathbf{q}}$ of the phonon vacuum deformation in the $\mathbf{q}$-th harmonics due to EPI are simply expressed through the corresponding Fourier-transform of the squared electronic wave function [33], $v_{\mathbf{q}}$ is a number of phonons in the $\mathbf{q}$-th harmonics radiated at the coherent state decay after the photoelectron escape [20,16].

The energy conservation law ($\delta$-function argument in Eq. (5)) relates the photoelectron energy and the number $v$ of radiated phonons: $E_p + \hbar\Omega = E_{kin} + \Phi + v\hbar\omega$, where $E_p, \hbar\Omega, E_{kin}$ and $\Phi$ are the polaron binding energy, photon energy, photoelectron energy and work function, respectively. Ordinarily in ARPES three latter values are replaced by binding energy $\varepsilon = E_{kin} - (\hbar\Omega - \Phi)$ so that the energy conservation law has the form $E_p = \varepsilon + v\hbar\omega$. It (and, consequently, the argument of $\delta$ – function in Eq.(5)) contains only two variables: $\varepsilon$ and $v$. Then in neglecting the phonon dispersion a probability $A(\mathbf{k},\varepsilon)$ to catch a photoelectron with the momentum $\mathbf{k}$ and binding energy $\varepsilon$, results from summarizing (5) over all the cases with equal total number $v$ of radiated phonons [16,20]:

$$A(\mathbf{k}, \varepsilon) \propto |\psi_{\mathbf{k}}|^2 P_{v(\varepsilon)}, \tag{7}$$

$$P_{v(\varepsilon)} = \sum_{\{v_{\mathbf{q}}\}=v} \prod_{\mathbf{q}} \left| \langle v_{\mathbf{q}} | d_{\mathbf{q}} \rangle \right|^2 = \frac{\bar{v}^{v(\varepsilon)-1}}{(v(\varepsilon)-1)!} e^{-\bar{v}}, \tag{8}$$

$$\nu(\varepsilon) = (E_p - \varepsilon)/\hbar\omega, \quad \overline{\nu} = \sum_{\mathbf{q}} |d_{\mathbf{q}}|^2 = E_{pvd}^{pol}/\hbar\omega, \tag{9}$$

where $\psi_{\mathbf{k}}$ is Fourier-transform of the electron wave function in the initial state, $\overline{\nu}$ is the average number of radiated phonons and $E_{pvd}^{pol}$ is the average polarization field energy in the polaron state due to the phonon vacuum deformation. According to known Pekar results [22] $E_{pvd}^{pol} = 2E_p$, and $E_p$ is determined by Eq.(3). Thus, only two system characteristics - $\varepsilon^*$ and $m^*$ - determine the ARPES spectrum according to Eqs.(7-9) at low doping and at higher doping while $n<n_0$ (as it will be shown below) through the AS binding energy, energy of the polarization field and electron wave function in AS. Polaron and bipolaron radius dependence on $\varepsilon^*$ and $m^*$ is depicted in [14]. At $n>n_0$ the carrier dispersion in the whole first BZ is involved into the calculation.

Due to phonon energy dispersion the energy dispersion curves $A(\mathbf{k}=const, \varepsilon)$ are envelopes of points obtained according to Eqs.(7-9). Such a calculation yields broad bands of Gaussian shape [16] located in the momentum region $k<k_0$ where $\psi_{\mathbf{k}}$ is noticeable. The calculated spectrum is demonstrated by Fig.2a and is in good accordance with the experimental ones [4,5] (of course, as we use the experimental band maximum to determine $\varepsilon^*$ the comparison at low doping makes sense for the width and shape of the calculated band in the energy and momentum space).

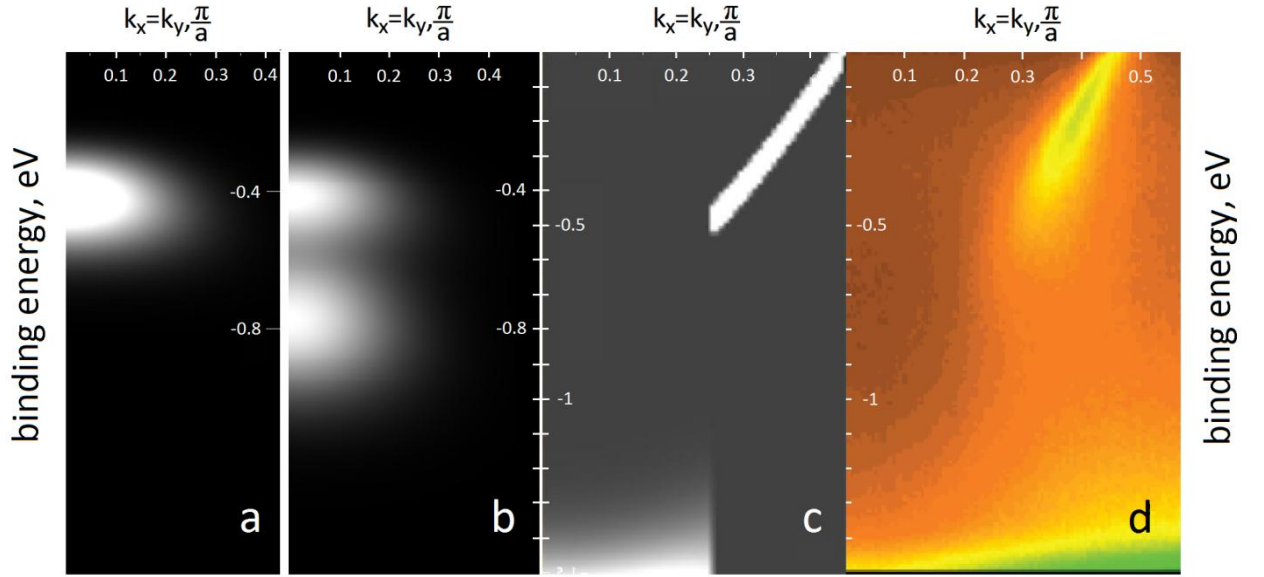

Fig.2 (color online) a,b,c - calculated ARPES spectra of electron-doped cuprate at low doping, at doping corresponding to coexistence of polarons and bipolarons and at electron concentration $n>n_0$, respectively; d – experimental spectrum along the nodal cut of $Nd_{1.83}Ce_{0.17}CuO_4$ [9].

Below we generalize the described method to systems with higher carrier concentration on the base of the following approach. Since the ground state degeneration occurs only at carrier concentration $n<n_0$, post-photoemission relaxation of a system with $n>n_0$ involves electron subsystem as well as phonon one. As quasiparticles (autolocalized and delocalized carriers) can be considered as non-interacting with each other the vector of the system state is a product of their vectors of state. Energy conservation equation describing the relaxation result includes only quasiparticles which change their states: the initial state of the photoelectron (localized at $k<k_0$ and delocalized at $k>k_0$) and the state with the maximum energy varied by doping.

5. Electron-doped cuprates case at increasing doping

Eqs.(5-9) were first generalized to describe photoemission from large bipolaron when its binding energy and wave function are calculated with taking into account carriers correlation inside polarization potential well [14]. However, the exact expression obtained was cumbersome.

To simplify it one can calculate the electronic matrix element neglecting carriers correlation in the bipolaron (electronic wave function in the bipolaron is approximated as a product of two polaron ones), but still taking it into account calculating bipolaron binding energy. Then $|\psi_\mathbf{k}|^2$ in (7) coincides with the polaron case and the difference of bipolaronic $A(\mathbf{k}, \varepsilon)$ from the polaronic one is in the energy conservation equation:

$$E_b = E_p + \varepsilon + \nu\hbar\omega \qquad (10)$$

where $E_b$ is the bipolaron binding energy per carrier. Eq.(10) determines $\nu(\varepsilon)$ to be substituted in (8). The average number of radiated phonons

$$\bar{\nu} = \Delta E_{pvd}/\hbar\omega, \quad \Delta E_{pvd} = E_{pvd}^{bip} - E_{pvd}^{pol}, \qquad (11)$$

where $\Delta E_{pvd}$ is the difference between the energies of the phonon vacuum deformation in the bipolaron and in the polaron.

The values $E_b$ and $E_{pvd}^{bip}$ for the one-center bipolaron with taking into account carriers correlation [14] are mainly determined by the same medium parameters $m^*$, $\varepsilon^*$, though depend slightly on $\varepsilon_\infty$. They can be calculated approximately with the mistake lower than 5% as (in eV):

$$E_b = -1.44(c^2 + \varepsilon^* cd + 1/80)m^*/m_e, \quad E_{pvd}^{bip} = 4*1.44(c/\varepsilon^* + d - 1/140)m^*/m_e, \qquad (12)$$

where $c = 2/\varepsilon^* - 1/\varepsilon_\infty$, $d = 2/(21\varepsilon_\infty^2)$. We use the value of $\varepsilon_\infty = 3$ [35] in all further calculations. The average number of phonons (11) radiated at the bipolaron polarization cloud decay down to polaron one is larger than (9) therefore the bipolaronic band in ARPES spectrum lies essentially deeper than the polaronic band. Fig.2b shows them both that corresponds to carrier concentration characteristic for coexistence of polarons and bipolarons (occurring in a narrow interval of the carrier concentration [14]). Similar smearing of the spectral weight over a wide area was observed in hole-doped cuprate at intermediate doping [35].

As it was pointed out above at the carrier concentration $n>n_0$ the system relaxation after photoemission involves electron subsystem due to absence of the ground state degeneration. At $n>n_0$, $k>k_0$ the relaxation does not affect the phonon vacuum. In this case two variants of the calculation yield the same result presented by Fig 2c: traditional one where the final state is a hole with the in-plane momentum k and one including relaxation of the hole to its minimum energy state on the Fermi surface. The energy conservation equation for the second calculation is $0 + E(k) = E(k) + \varepsilon + \nu\hbar\omega$, where $E(k)= E_{UHB}(k)$ is determined by Eq.(1) with zero of energy at the Fermi surface (changing with the doping), $\nu = (0 - E(k))/\hbar\omega$ is the number of radiated phonons. This yields $\varepsilon(k)=E(k)$, i.e. spectral weight in the region $k>k_0$ follows "bare" electron dispersion crossing the Fermi level. This result allows using the electron band dispersion obtained from the experimental ARPES spectrum at sufficiently high doping for the carrier effective mass calculation.

At $n>n_0$, $k<k_0$ the initial state is bipolaron. Therefore $|\psi_\mathbf{k}|^2$ in (7) results in the same localization of the bipolaron band in the momentum space as at $n<n_0$. However, the other relaxation way changes its position in the binding energies. Similarly to $n<n_0$ case after photoelectron escape the bipolaron polarization cloud begins to decay down to the polaron one with radiation of phonons, their average number is $\bar{\nu}_1 = \Delta E_{pvd}'/\hbar\omega$, provided the process is completed. The stroke designates that at $n>n_0$ the difference between the polarization field energy in the bipolaron and in the polaron is larger than in the single bipolaron case due to ASs contraction in the system ground state at high carrier density. Indeed, Eqs.(12) correspond to the single bipolaron case. At $n>n_0$ we consider the case of the highest possible density of bipolarons. The ground state of the system in such a case corresponds to "compressed" bipolaron liquid. For the case under consideration where $(\varepsilon^*)^{-1}=0.3$ the bipolaron radius decreases about 2 times due to their compression in the ground state at high carrier concentration. Accordingly to Eq.(4) simultaneous increase of $k_0$ value occur. Bipolarons contraction results also in their binding energy decrease whereas the polarization field energy grows.

At $n>n_0$ the relaxation due to strong EPI restores the bipolaron after photoelectron with $k<k_0$ escape. First a transition of a delocalized carrier towards the minimum energy state (or relaxation of the photohole) occurs. This process and decay of the bipolaron polarization "coat" into polaron one are both multi-phonon processes and occur during the characteristic phonon times. Therefore there will be a distribution of probabilities for the degree of the latter process completeness. As a result some decrease of $\bar{v}_1$ will take place. However, this decrease is compensated to some degree by increase of $\Delta E_{pvd}'$ due to bipolarons contraction at high carrier density. Therefore for rough estimate of energetic position of the spectral weight at $k<k_0$ we suppose $\bar{v}_1 \approx \Delta E_{pvd}/\hbar\omega$. The possible mistake is not essential in comparison with extremely large width of the ARPES band at $k<k_0$.

Relaxation empties the Fermi-surface state, its excess energy and momentum are transferred to the phonon field with creating the appropriate average number $\bar{v}' = \hbar^2 k_F^2/(2m^*\hbar\omega)$ of phonons. The bipolaron restoration from the polaron is also accompanied with the radiation of phonons, their average number $\bar{v}_2 \approx \bar{v}_1$ since the initial and final phonon vacuums for the cases of decay and restoration are simply interchanged. According to Gaussian distribution property the average number of radiated phonons

$$\bar{v} = \bar{v}_1 + \bar{v}_2 + \bar{v}' \cong 2\bar{v}_1 + \bar{v}' \cong 2\Delta E_{pvd}/\hbar\omega + \hbar^2 k_F^2/(2m^*\hbar\omega). \tag{13}$$

The energy conservation equation at $n>n_0$, $k<k_0$ has the form $E_b + 0 = E_b + \varepsilon + v\hbar\omega$. Using it to express $v(\varepsilon)$ for substitution into $P_v$ (8) simultaneously with $\bar{v}$ value (13) we calculate spectrum at $k<k_0$ according to Eq.(7). The result is demonstrated by Fig.2c. It is in good *quantitative* agreement with the experimental spectrum [9] shown on Fig.2d, both spectra are in one and the same scale.

The theoretical calculation of the case of optimal doping with electrons in the frames of Hubbard model without EPI yields [9] the low-energy part of the ARPES band (caused by delocalized carriers) in good accordance with the experiment. It is natural since this part follows the "bare" carrier dispersion as it is shown above. The most noticeable difference between the spectra calculated in the present approach from those obtained in models without EPI [9,10] is the broad band of Gaussian shape observed well below the Fermi energy in the restricted area of k-space in ARPES spectrum at low doping [4,5] which appears in models with strong EPI according to Eqs. (7-9) and does not arise in models without EPI.

6. Undoped and hole-doped cuprates cases

To calculate photoemission from undoped and hole-doped cuprates we use the correlated carriers dispersion (2) modified by strong Frohlich EPI (Fig.1b). Similarly to the electron-doped systems with $n>n_0$ there are two types of the initial state: autolocalized (bipolaron) for photoelectrons with the momentums $k<k_0$ and delocalized for $k>k_0$. The relaxation process is also similar to high ($n>n_0$) electron doping case with the only difference in the final state. The system states with delocalized hole or electron polaron which are created at the photoemission from the states $k>k_0$ and $k<k_0$, respectively, are not its stationary state. Strong EPI results in hole polaron formation (with preceding hole transition into a state near the electron band maximum). Deformation of the phonon vacuum during the hole polaron formation is accompanied by multiple phonon radiation. As a result the whole ARPES spectrum both in $k<k_0$ and $k>k_0$ regions is shifted deeper in the binding energy by about $3E_p^h$ (where $E_p^h$ is the hole polaron binding energy) and in the region $k>k_0$ the ARPES band is broad analogously to polaronic band in electron-doped system.

At $k<k_0$ the energy conservation equation and the average number of radiated phonons are

$$E_b + 0 = E_b + \varepsilon + E_p^h + v\hbar\omega, \quad \bar{v} \cong (2\Delta E_{pvd} + 2E_p^h - E_{LHB}(k))/\hbar\omega, \quad (14)$$

where $E_{LHB}(k)$ is determined by Eq.(2) with zero energy at the Fermi level (changing with doping). Thus, in undoped cuprates the ARPES band in the region $k<k_0$ lies deeper than in the electron doped ones at high doping by the energy released at hole polaron formation and preceding hole relaxation towards its minimum energy state. At $k>k_0$

$$E_{LHB}(k) + 0 = \varepsilon + E_{LHB}(k) + E_p^h + v\hbar\omega, \quad \bar{v} \cong (2E_p^h - E_{LHB}(k))/\hbar\omega, \quad (15)$$

The ARPES band dispersion at $k>k_0$ obtained as average of the first Eq.(15) has the form

$$\bar{\varepsilon}(k) = -E_p^h - \bar{v}\hbar\omega = -3E_p^h + E_{LHB}(k), \quad (16)$$

i.e. it follows the "bare" carrier dispersion shifted deeper by $3E_p^h$. This is similar to the SP model [12,13] result but in SP model this dispersion is predicted for any $k$ value whereas in the LP model the "vertical dispersion" emerges at $k\approx k_0$ due to different system relaxation ways in different $k$ regions. Fig.3a,b demonstrate calculated and experimental [11] ARPES spectrum from undoped cuprate in one and the same scale.

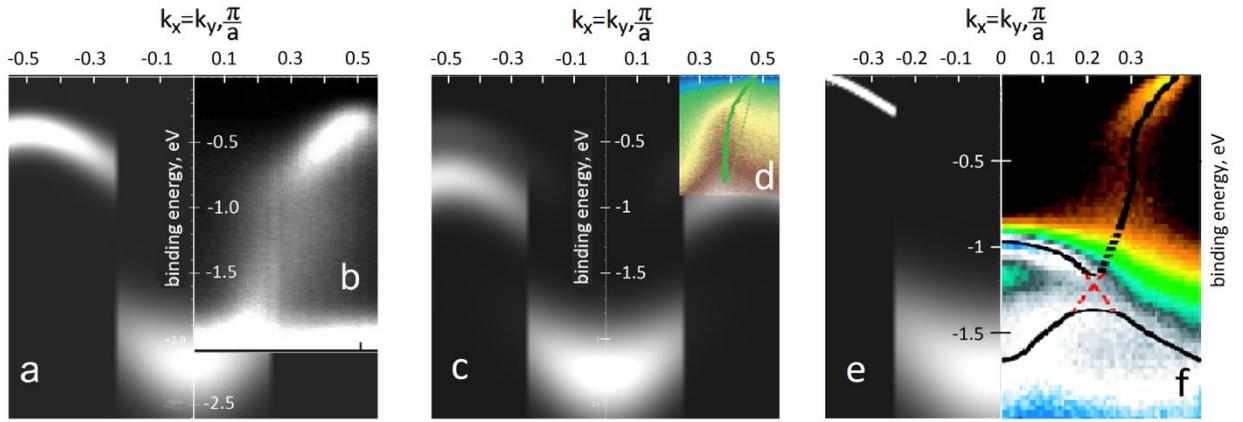

Fig.3 (color online) a,b - calculated and experimental (along nodal cut of $Ca_2CuO_2Cl_2$) [11] ARPES spectra of undoped parent compound, respectively, in the same scale; c – calculated ARPES spectrum at hole doping corresponding to two possible final states: hole polaron and bipolaron; d – experimental spectrum along nodal cut of highly underdoped $Bi_2Sr_2CaCu_2O_{8+\delta}$ ($T_c$=5K) [35] in the same scale as panel c; e - calculated ARPES spectrum at hole concentration $n>n_0$ (intensity map); f – experimental ARPES spectrum of highly overdoped Bi2201 in nodal direction [6] in the same scale as panel e.

Fig.4 represents the calculated energy distribution curves (EDCs) obtained from the spectral function shown by Fig.3a at constant values of $k\geq k_0$. The shape of the calculated band is Gaussian, its width (determined by $\bar{v}$ value (15)) increases at moving away from the band extremum in accordance with the experiment [3]. In experimental ARPES spectra this effect was highlighted in [3] where the ratio of the binding energy in the EDC maximum to the EDC halfwidth (width at the half height) was obtained approximately constant and equal to 1.8. In the calculated EDCs presented in Fig.4 this ratio depends on the phonon energy and for the average phonon energy 0.04 eV it is 1.8 at $k_x = k_y = 0.5\pi/a$ (in the ARPES band maximum) and 1.9 at $k_x = k_y = 0.25\ \pi/a$ where (approximately) the "vertical dispersion" emerges in good agreement with the experiment [3].

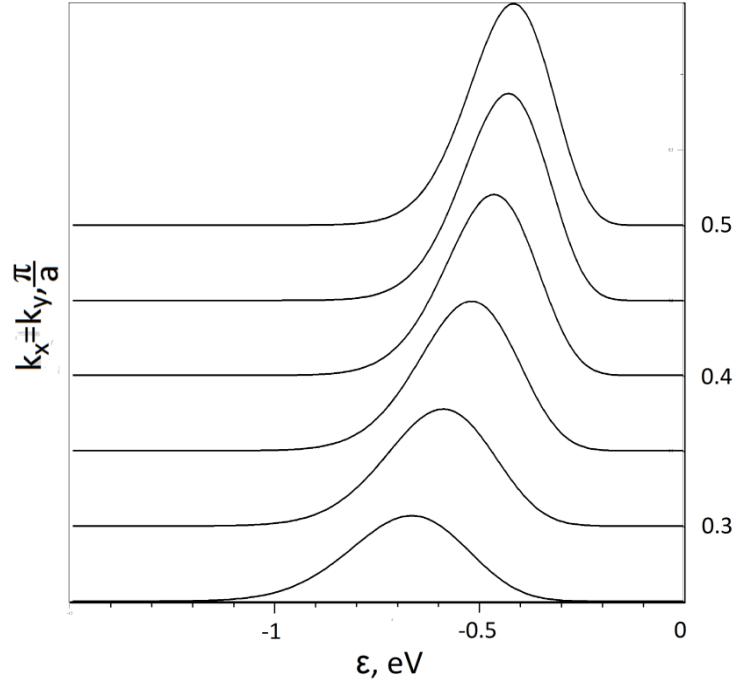

Fig.4. Energy distribution curves calculated for undoped system with the nodal dispersion (2) and $(\varepsilon^*)^{-1}=0.3$.

When at increasing doping the hole polaron states become filled the final photohole state is no longer a hole polaron but a hole bipolaron. The energy conservation equations read

$$E_b + E_p^h = E_b + \varepsilon + E_b^h + v\hbar\omega, \quad \bar{v} \cong (2\Delta E_{pvd} + \Delta E_{pvd}^h - E_{LHB}(k))/\hbar\omega, \quad k<k_0, \qquad (17)$$

$$E_{LHB}(k) + E_p^h = \varepsilon + E_{LHB}(k) + E_b^h + v\hbar\omega, \quad \bar{v} \cong (\Delta E_{pvd}^h - E_{LHB}(k))/\hbar\omega, \quad k>k_0, \qquad (18)$$

where $E_{LHB}(k)$ is determined by Eq.(2) with zero of energy at the Fermi level. Similarly to the electron doping system the spectral weight in the case of bipolaron final state is shifted deeper in comparison with the polaron one since the energy $\Delta E_{pvd}^h + E_b^h - E_p^h$ (where h denotes hole ASs) is larger than $3E_p^h$. Its location in the momentum space does not change.

If due to close binding energy per carrier in the polaron and bipolaron state the photohole can appear in some concentration interval with comparative probability in both these states the spectral weight will be smeared over more wide area of energies. Interestingly, that this stage of the spectral weight transfering deeper upon doping the authors of [35] managed to observe on the very underdoped cuprate with $T_c=5K$. Calculated and experimental ARPES spectra demonstrating smearing the spectral weight over very wide area of energies are shown on Fig.3c,d, respectively, in one and the same scale. Smaller intensity of the polaronic band in comparison with the bipolaronic one in experimental spectrum (Fig.3d) likely means that the doping level of the pattern correponds to almost filled hole polaron states, the same situation is chosen for calculated spectrum shown on Fig.3c.

At hole concentration $n>n_0$ the final state of the photohole is delocalized state at the Fermi surface. Since it does not induce the phonon vacuum deformation the whole HEP of the ARPES spectrum is lifted (in comparison with the $n<n_0$ case) along the energy axis by the value $3.2E_p^h + |E(\pi/2a) - E(k_F)|$. Energy conservation equations have the form

$$E_b + 0 = E_b + \varepsilon + v\hbar\omega, \quad \bar{v} \cong (2\Delta E_{pvd} - E_{LHB}(k))/\hbar\omega, \quad k<k_0, \qquad (19)$$

$$E_{LHB}(k) + 0 = \varepsilon + E_{LHB}(k) + v\hbar\omega, \quad v = -E_{LHB}(k)/\hbar\omega, \quad k>k_0. \qquad (20)$$

The ARPES band dispersion at $k>k_0$ follows unshifted "bare" carrier dispersion similarly to the electron doping case at $n>n_0$.

The HEP of the ARPES spectrum calculated according to Eqs.(7,8,19, 20) is presented in Fig.3e. Fig.3f shows the experimental spectrum of highly overdoped Bi2201 in nodal direction [6] in the same scale. The comparison is complicated due to crossing of several bands in the experimental spectrum. Similar dispersion was also reported in [7], the similar dispersion and band crossing were observed in [8]. Taking into account that the LHB dispersion can be slightly different in $Ca_2CuO_2Cl_2$ (it was taken from [11] and used to write Eq.(2)) and in Bi2201 and Pb-Bi2212 studied in [6-8] the calculated HEP of the ARPES spectrum is in good agreement with experimental results.

7. Conclusion

In summary, we suggest a new approach to calculating the HEP of ARPES spectrum from cuprates which takes into account joint relaxation of strongly coupled fields: a field of correlated electrons and phonon field. Such type of the relaxation occurs only in the case of strong long-range EPI at sufficiently high carrier concentration when degeneration of the ground state is absent due to coexistence of autolocalized and delocalized carriers. Separation of the momentum space between two these types of carrier states according to Pauli exclusion rule together with different relaxation pathways for autolocalized and delocalized initial states result in fragmentation of the band in ARPES spectrum into two parts in different regions of the wave vector space. Predicted theoretically value of the "vertical dispersion" momentum position (the $k_0$ value) is in good agreement with the experiments on cuprates with both types of doping. The phonon field relaxation at formation or decay of the carrier ASs causes broad bands of Gaussian shape in the ARPES spectrum.

Besides realizing the possible reason of fragmentation of the HEP in ARPES spectrum of cuprates we suggest a new method to calculate analyticaly the HEP of ARPES spectrum in systems with strong carrier correlations and strong long-range EPI at arbitrary carrier concentration. It takes advantages of applying coherent states basis for the phonon field state description. As the input the method uses dispersion of the correlated carriers and the effective dielectric constant characterising the EPI. The calculated position, width and shape of the bands in the HEP of ARPES spectrum of cuprates are in good agreement with the experiments at any level of doping with both types of carriers.

The experimentally observed double scale of HEA energy in electron-doped cuprates with respect to that in hole-doped cuprates arises naturally in the frames of model under consideration. Indeed, in this approach the momentum $k_0$ of the break in the delocalized carrier dispersion at essential doping due to Pauli exclusion rule is invariant in electron and hole-doped cuprates due to close size of the bipolarons in them. But the corresponding energy $E(k_0)$ (the energy of HEA) is determined by the LHB and UHB dispersion which are different as it is illustrated for nodal direction by Eqs.(1) and (2) and Fig.1a,b.

Thus, the suggested approach allows to obtain theoretically both characteristic features observed in HEP of experimental ARPES spectra of cuprates: broad bands of Gaussian shape and "vertical dispersion" patterns. This differs it from the previous theoretical results obtained in models considering electron correlations without taking into account strong EPI [7,9,10] (where broad bands of Gaussian shape observed experimentally at zero and low doping [2-5,11] do not appear) or allowing for strong short-range EPI [12,13] (where the "vertical dispersion" patterns are absent).

The present approach allows also understanding several experimental findings being unclear in some earlier models. First, manganites are also characterized by broad bands in ARPES spectrum but do not demonstrate "vertical dispersion" patterns that was considered as proof that strong EPI cannot be the reason of this feature [36]. As we have shown strong long-range EPI causes "vertical dispersion" at sufficiently high doping due to coexistence of autolocalized anf delocalized carriers, whereas strong short-range EPI resulting in small polaron formation does not. However both types of EPI result in broad Gaussian bands in ARPES spectrum [12,13,16] provided the EPI is strong. Thus, the mentioned difference in ARPES spectra of cuprates and manganites is likely caused by different dominating type of EPI, long-range in the former case and short-range in the latter, albeit EPI is strong in both cases.

Second, some notes concerning the observation conditions of HEA can be made. They are important because changes in HEA manifestation in some cuts in different BZs at change of the photon energy observed experimentally [25-27] posed a question whether it isn't an effect of the matrix elements only. We show that the break of the delocalized carrier dispersion in essentially doped systems due to strong EPI exists in restricted region of the k space: at $|\mathbf{k}| < k_0$. Thus, two of three cuts studied in [25] are out of the region where HEA occurs and the third cut $k_x = 3\pi/(8a)$ is just on the boundary $|\mathbf{k}| = k_0$. Likely, this is the reason of unordinary behavior of the spectrum [25] taken from this cut at the photon energy change. Earlier the similar behavior was observed in the second and third BZs [26,27]. However, if the HEA nature is related with the strong long-range EPI breaking the translational symmetry due to ASs formation then the appearance of HEA in the second and third BZs in the same form as in the first one is questionable.

Finally, demonstrated broken by strong EPI translational symmetry in cuprates allows also discussion of charge ordering observed in doped cuprates [28-31] in terms of the large-radius polarons and bipolarons formation. Indeed, the estimated radius of the bipolaron at their maximum density in cuprates is $R_{bip} \approx 6.5 \div 7$ Å as was mentioned above. This value is in good agreement both with the $k_0$ value marking the HEA position in the momentum space [6-11] and with the period of charge ordering ($3.3 \div 4a$) [28-31]. Of course, this is a subject of separate consideration as we used here the simplest quasi-isotropic model whereas to describe the square grid of ordered charges [28-31] one should take into account the dispersion anisotropy.

The agreement of the calculated HEP of the ARPES spectrum with the experiments on cuprates confirms coexistence of autolocalized and delocalized carriers in them. This result may be useful for understanding temperature and doping dependence of their transport and magnetic properties [23]. Suggested approach may be also effective in theoretical modeling the evolution of the optical conductivity spectra of cuprates with doping and temperature [37,38].